\documentclass[aps,preprint,showpacs]{revtex4}
\usepackage{epsfig}
\usepackage{amsmath}

\begin{document}

\title{A thermodynamical fiber bundle model for the fracture of disordered
materials}
\author{Alessandro Virgilii and  Alberto Petri \\
 Consiglio Nazionale delle Ricerche, Istituto dei Sistemi Complessi, \\
Via del Fosso del Cavaliere 100,
00133, Roma, Italy \\ and  \\ Silvio R. Salinas \\
Instituto de F\'{\i}sica, Universidade de S\~{a}o Paulo,\\ Caixa
Postal 66318,  05315-970 S\~{a}o Paulo, SP, Brazil}
\date{\today }

\begin{abstract}
We investigate a disordered version of a thermodynamic fiber
bundle model proposed by Selinger, Wang, Gelbart, and Ben-Shaul a
few years ago.  For simple forms of disorder, the model is
analytically tractable and displays some new features. At either
constant stress or constant strain, there is a non monotonic
increase of the fraction of broken fibers as a function of
temperature. Moreover, the same values of some macroscopic
quantities as stress and strain may correspond to different
microscopic configurations, which can be essential for determining
the thermal activation time of the fracture. We argue that
different microscopic states may be characterized by an
experimentally accessible analog of the Edwards-Anderson
parameter. At zero temperature, we recover the behavior of the
irreversible fiber bundle model.
\end{abstract}

 \maketitle
\section{Introduction}

It is known that some features of the phenomena of fracture can
still be captured by simple and schematic models
\cite{roux90,chakrabarti97,alava05}. The simple Fiber Bundle (FB)
model, introduced by Peirce \cite{peirce26} and Daniels
\cite{daniels45} to account for the breakdown load in textiles, is
a simple model that has provided a useful general scheme for
describing the fracture of disordered systems in the quasi-static
regime. It indicates the critical onset \cite{hansen94} of the
fracturing process \cite {petri94,caldarelli96}, and shares a
number of features with other models, including a correspondence
with results from a mean-field approach \cite {zapperi99} that is
very effective to deal with elasticity
\cite{chakrabarti97,alava05}.

Macroscopic fracture is a global irreversible process. At the
microscopic level, however, microcracks can heal under suitable
conditions, which gives room to the consideration of reversible or
at least partially reversible models. In delayed (activated)
fractures, the failure load may decrease as we increase the
temperature or the time of loading \cite{shahidzadeh05}. Delayed
fracture phenomena have been investigated by using both fully and
partially reversible models
\cite{golubovich91,pauchard93,kitamura97,kitamura01,bonn98,politi02,
santucci03,santucci04,sornette05}.
In this article, we study the behavior of an extension of a
reversible thermodynamical model that has been proposed a few
years ago by Selinger, Wang, Gelbart, and Ben-Shaul
\cite{selinger91}. The model consists of a bundle of independent
and identical fibers (elastically harmonic strings, with the
addition of a failure energy level), which are loaded in parallel,
and whose states, broken or intact, are allowed to fluctuate
according to (reversible) Boltzmann weights. Despite its enormous
simplicity, this model still captures some relevant and essential
features of homogeneous nucleation of fractures in defect-free
crystals, with relevant predictions for systems as iron whiskers
\cite{brenner62}, which remain intact, in a metastable state,
until eventually undergoing a fracture.

It has been recently pointed out that delayed fracture can also
occur in non homogeneous materials
\cite{roux90,santucci04,shahidzadeh05}. We then decided to
investigate the Thermodynamic Fiber Bundle (TFB) model of Selinger
and coworkers in the presence of disorder. Although we recognize
that the full description of failure requires the introduction of
some degree of irreversibility, even at finite temperatures, we
believe that the present investigation is a first step towards an
understanding of the underlying processes of thermally activated
fractures. In several simple cases, the general thermodynamic
properties of this Disordered Thermodynamic Fiber Bundle (DTFB)
model are analytically accessible, and may differ from the
corresponding results for the FB and the homogeneous TFB models.
In particular, depending on the type of disorder, the fraction of
intact fibers can display a non-monotonic behavior with
temperature and constant applied strain (or stress). The same
macroscopic load (strain, stress) thus corresponds to different
microscopic configurations, and this  may have an important
influence on the activation time of the fracture. We conjecture
that these configurations may be distinguished by an analog of the
Edwards-Anderson parameter, and that the value of this parameter
can be measured from macroscopic quantities.

In Sec. II we introduce the TFB model, and recover the original
results of Selinger and coworkers. In Sect. III, we define the
disordered DTFB model, and analyze some realizations of disorder
at fixed strain. The experimentally more relevant case of fixed
stress is discussed in Sect. IV. Section V is devoted to the
discussion of the Edwards-Anderson parameter, whereas some final
comments and a summary of the results are given in Section VI.

\section{The TFB model}

The homogeneous Thermodynamic Fiber Bundle (TFB) model consists of
$N$ fibers (harmonic springs) subjected to the same load. Let the
set of variables $ \left\{ t_{i};i=1,...N\right\} $ describe a
configuration with broken ($ t_{i}=0$) and intact ($t_{i}=1$)
fibers. The energy (Hamiltonian) of a configuration of this system
is given by
\begin{equation}
H=\frac{1}{2}k\varepsilon ^{2}\sum_{i=1}^{N}t_{i}-D\sum_{i=1}^{N}t_{i},
\label{s1}
\end{equation}
where $k>0$ is an elastic modulus, $\varepsilon $ is the common
strain, and $ D>0$ is the (uniform) dissociation energy. We assume
that the breaking process is reversible, with probabilities
related to the Boltzmann factors, and write a canonical partition
function in the \textquotedblleft strain\textquotedblright\
ensemble,
\begin{equation*}
Z=\sum_{\left\{ t_{i}\right\} }\exp \left[ -\frac{1}{2}\beta k\varepsilon
^{2}\sum_{i=1}^{N}t_{i}+\beta D\sum_{i=1}^{N}t_{i}\right] =
\end{equation*}
\begin{equation}
=\left[ 1+\exp \left( -\frac{1}{2}\beta k\varepsilon ^{2}+\beta
D\right) \right] ^{N},  \label{s3}
\end{equation}
where $\beta =1/T$ is the inverse of the temperature. From this
partition function, we have the Helmholtz free energy per fiber,
\begin{equation}
f\left( T,\varepsilon \right) = - \frac{1}{\beta} \ln \left[
1+\exp \left( -\frac{1}{2}\beta k\varepsilon ^{2}+\beta D\right)
\right] . \label{helm}
\end{equation}
Given a configuration $\left\{ t_{i}\right\} $, the fraction of intact
fibers is written as
\begin{equation}
\phi =\frac{1}{N}\sum_{i=1}^{N}t_{i},  \label{s2}
\end{equation}
which leads to the canonical average
\begin{equation}
\label{TFB} \left\langle \phi \right\rangle_T =\frac{1}{1+\exp
\left( \beta k\varepsilon ^{2}/2-\beta D\right) },
\end{equation}
as in the original work of Selinger et al.

In order to make contact with the calculations of Selinger and coworkers, we
can rewrite the canonical partition function as an integral over the
fraction of intact fibers $\phi $,
\begin{equation}
Z=\sum_{\left\{ t_{i}\right\} }\int d\phi \delta \left( \phi
-\frac{1}{N} \sum_{i=1}^{N}t_{i}\right) \exp \left\{ \beta N\left[
-\frac{1}{2} k\varepsilon ^{2}\phi + D \phi \right] \right\} .
\end{equation}
Using an integral representation for the Dirac $\delta $-function, and
discarding irrelevant terms in the thermodynamic limit (large $N$), it is
easy to see that
\begin{equation}
Z=\int d\phi \exp \left[ -\beta N\mathcal{F}\right] ,
\end{equation}
where
\begin{equation}
\mathcal{F}=(\frac{1}{2}k\varepsilon ^{2}-D)\phi +\frac{1}{\beta
}\left[ \left( 1-\phi \right) \ln \left( 1-\phi \right) +\phi \ln
\phi \right] .
\end{equation}
The minimization of $\mathcal{F}$ with respect to $\phi $
leads to
the Helmholtz free energy of Eq. (\ref{helm}).

We now consider the more realistic situation of fixed stress $\sigma $. In
the stress ensemble, we write the partition function
\begin{equation*}
Y=\int d\varepsilon \exp \left( -\beta \sigma \varepsilon N\right)
Z=\sum_{\left\{ t_{i}\right\} }\exp \left[ \beta
D\sum_{i=1}^{N}t_{i}+\frac{ \beta N\sigma
^{2}}{2k\sum_{i=1}^{N}t_{i}}\right] =
\end{equation*}
\begin{equation}
=\sum_{\left\{ t_{i}\right\} }\int d\phi \delta \left( \phi
-\frac{1}{N} \sum_{i=1}^{N}t_{i}\right) \exp \left\{ \beta N\left[
D\phi +\frac{\sigma ^{2}}{2k\phi }\right] \right\} .  \label{s5}
\end{equation}
Again, if we use an integral representation of the $\delta $-function, and
discard irrelevant terms, it is easy to write
\begin{equation*}
Y=\int d\phi \exp \left[ -\beta N\mathcal{G}\right] ,
\end{equation*}
with
\begin{equation}
\mathcal{G}=-\frac{\sigma ^{2}}{2k\phi }-D\phi +\frac{1}{\beta
}\left[ \left( 1-\phi \right) \ln \left( 1-\phi \right) +\phi \ln
\phi \right] ,
\end{equation}
whose minimum with respect to $\phi $ leads to the expression of
the Gibbs free energy per fiber bond. Selinger and coworkers point
out that the function $\mathcal{G}$ in terms of $\phi $, for
sufficiently small values of the stress $\sigma $, displays a
local relative minimum at $\phi ^{\ast }$, with $0<\phi ^{\ast
}\leq 1$, which corresponds to a metastable state with a fraction
$\phi ^{\ast }$ of intact fibers. Also, it is shown that
increasing the stress $\sigma $, there is a decrease of both the
values of $\phi ^{\ast }$ and of the height of the barrier
separating the local minimum from the global minimum of the free
energy at $\phi =0$, with a fracture threshold at the spinodal
$\sigma _{c}=\sigma _{c}\left( T,D,k\right) $.

\section{The disordered DTFB model}

We now assume that each fiber bond is characterized by a different
dissociation energy $D_{i}$, for $i=1,...N$. The disorder is
fixed, quenched, so that $\left\{ D_{i}\right\} $ is a given set
of independent and identically distributed, random variables, with
a probability distribution $p\left( D_{i}\right) $. The energy of
a configuration of this system is given by
\begin{equation}
H=\frac{1}{2}k\varepsilon ^{2}\sum_{i=1}^{N}t_{i}-\sum_{i=1}^{N}D_{i}t_{i}.
\label{s9}
\end{equation}

Again, we assume a thermodynamic process, at temperature $T$, with Boltzmann
weights. In the canonical strain ensemble, given the configuration $\left\{
D_{i}\right\} $, we write the partition function
\begin{equation}
Z\left\{ D_{i}\right\} =\prod_{i=1}^{N}\left[ 1+\exp \left(
-\frac{1}{2} \beta k\varepsilon ^{2}+\beta D_{i}\right) \right].
\end{equation}
In the thermodynamic limit, we can use the law of large numbers in order to
write the (self-averaging) Helmholtz free energy
\begin{equation}
f\left( T,\varepsilon \right) =-\frac{1}{\beta }\left\langle \ln \left[
1+\exp \left( -\frac{1}{2}\beta k\varepsilon ^{2}+\beta D\right) \right]
\right\rangle ,  \label{average}
\end{equation}
where
\begin{equation}
\left\langle \cdot \cdot \cdot \right\rangle =\int dDp\left( D\right) \cdot
\cdot \cdot .
\end{equation}

The average fraction of integer fibers is given by
\begin{equation}
\phi (\epsilon ,T)=\left\langle \frac{1}{1+\exp \left( \frac{1}{2}\beta
k\varepsilon ^{2}-\beta D\right) }\right\rangle ,  \label{genphi}
\end{equation}
which resembles the form of the density of states of free fermions. In the
zero temperature limit, this fraction becomes a step function, so that
\begin{equation}
\label{cumulate} \lim_{T\rightarrow 0}\phi
(\epsilon,T)=1-P(\frac{1}{2}k\epsilon^2),
\end{equation}
where $P(D)$ is the cumulative probability distribution,
\begin{equation}
\label{gencumulate} P(D)\equiv \int_{0}^{D}dD^{\prime }p(D^{\prime
}),
\end{equation}
in agreement with the predictions for the irreversible FB model at constant
strain.

In the following subsections, we consider a situation of fixed strain, with
uniform and bimodal distributions of the dissociation parameter $D$.

\subsection{Uniform disorder}

In the usual version of the FB model \cite{chakrabarti97}, one
assumes a uniform distribution of dissociation energies and
identical harmonic elastic terms.  We then compare the main
features of the disordered DTFB model with the corresponding
results for the uniform TFB and the FB models. In the strain
ensemble, it is straightforward to calculate the averages given by
Eqs. (\ref{average}) to (\ref{gencumulate}). In particular, we
look at the average fraction of intact fibers, and the
stress-strain ($\sigma $ versus $\varepsilon $) characteristic
curve. If the dissociation energies are uniformly distributed over
the interval $[a,b]$, with $0<a<b$, we have
\begin{equation}
\phi =\frac{T}{b-a}\ln \left[ \frac{1+e^{-(\frac{1}{2}k\epsilon
^{2}-b)/T}}{ 1+e^{-(\frac{1}{2}k\epsilon ^{2}-a)/T}}\right] .
\end{equation}
In Figs. 1a) and 1b), we plot $\phi $ as a function of temperature
for different values of the strain. The corresponding behavior of
the uniform TFB model is shown in Figs. 1c) and 1d). In Fig. (2),
we plot $\phi $ as a function of strain, $\varepsilon$, for
different values of temperature $T$. The main differences are at
low temperatures. In this regime, the presence of disorder smooths
the transition from $\phi =1$, at low strain, to $\phi =0$, at
large strain. At $T=0$ the transition is abrupt in the absence of
disorder, but it becomes continuous in the disordered DTFB model.
In this case the curve is directly related to the probability
distribution for the dissociation energies, Eq.~(\ref{cumulate}),
and reproduces the behavior of the usual FB model, with a
discontinuity in the derivative. At low but non vanishing
temperatures, there are no singularities in both models, but in
the TFB model $\phi $ drops more quickly with increasing strain.
This is reflected in the stress-strain characteristic curve
$\sigma = k \phi \epsilon$ shown in Fig. 3a).

\subsection{Bimodal distribution of disorder}

Although the case of uniform disorder does not differ too much
from the case with no disorder, there are strong differences with
other forms of the distribution of disorder. For example, we may
consider a bimodal distribution,
\begin{equation}
\label{bimodal}
 p(D)=(1-c)\delta (D-D_{1})+c\delta (D-D_{2}).
\end{equation}
where $c$, $D_{1}$, and $D_{2}>D_{1}$ are non negative parameters.
In the strain ensemble, the Helmholtz free energy is thus given by
\begin{equation*}
f=-\frac{1}{\beta }\left( 1-c\right) \ln \left[ 1+\exp \left(
-\frac{1}{2} \beta k\varepsilon ^{2}+\beta D_{1}\right) \right] -
\end{equation*}
\begin{equation}
-\frac{1}{\beta }c\ln \left[ 1+\exp \left( -\frac{1}{2}\beta k\varepsilon
^{2}+\beta D_{2}\right) \right] ,
\end{equation}
with the average fraction of intact fibers
\begin{equation}
\phi =\frac{1-c}{1+\exp \left( \frac{1}{2}\beta k\varepsilon ^{2}-\beta
D_{1}\right) }+\frac{c}{1+\exp \left( \frac{1}{2}\beta k\varepsilon
^{2}-\beta D_{2}\right) }.
\end{equation}
In Figs. 4a) and 4b), we draw graphs of $\phi $ versus temperature
$T$ for different values of the strain, with $c=0.7$, $D_{1}=0$
and $D_{2}=1$. In contrast to the case of uniform disorder, the
graph of the average fraction $\phi $ displays pronounced extremal
points, which give rise to a counter-intuitive behavior. At small
values of the strain, the average fraction of integer fibers first
increases with temperature, and then drops slowly. At large values
of the strain, Fig. 4b), it first decreases and then increases
monotonically at higher temperatures. This behavior can be
understood if we consider that in this case $\phi $ behaves as a
linear combination of two TFBs with no disorder and different
dissociation energies. In each TFB $\phi $ is increasing for the
strain $ \varepsilon $ less than a certain threshold $\varepsilon
_{0}$, and decreasing for $\varepsilon >\varepsilon _{0}$, with
$\varepsilon _{0}=\left( 2D/k\right) ^{1/2}$, Eq. (\ref{TFB}).
With bimodal distribution, we anticipate that $\phi $ should
display a more or less pronounced extremum for  $\varepsilon
_{1}<\varepsilon <\varepsilon _{2}$ (with $\varepsilon _{i}=\left(
2D_i/k\right) ^{1/2})$, since in this region there is a
superposition of monotone increasing and decreasing curves. In
fact, the form of the disorder distribution is crucial for
determining the mechanical properties of the system. In
particular, characteristic load curves may differ widely,
depending on whether the disorder distribution has a finite
support, short tails, long tails, and other geometric features. A
study of this dependence will be the subject of a forthcoming
paper \cite{next}.

\section{The DTFB model in the stress ensemble}

In most of the real experimental situations, the external stress
$\sigma $ is given, instead of the internal strain $\varepsilon $.
As in Section III, given a disorder configuration $\left\{
D_{i}\right\} $, we write the partition function in the stress
ensemble,
\begin{equation*}
Y\left\{ D_{i}\right\} =\int d\varepsilon \exp \left( -\beta \sigma
\varepsilon N\right) Z\left\{ D_{i}\right\} =
\end{equation*}
\begin{equation}
=\sum_{\left\{ t_{i}\right\} }\exp \left[ \beta
\sum_{i=1}^{N}D_{i}t_{i}+ \frac{\beta N\sigma
^{2}}{2k\sum_{i=1}^{N}t_{i}}\right] .  \label{s11}
\end{equation}

In order to carry out the calculations it is useful to introduce a
new variable
\begin{equation}
Q=\frac{1}{N}\sum_{i=1}^{N}D_{i}t_{i},  \label{s12}
\end{equation}
which is a typical variable for disordered systems, representing a
sort of disorder-averaged magnetization, the projection of the
system configuration ${t_i}$ upon the disorder. Using $Q$ and the
definition of $\phi $, given by equation (\ref{s2}), we write
\begin{equation*}
Y\left\{ D_{i}\right\} =\sum_{\left\{ t_{i}\right\} }\int d\phi
\delta \left( \phi -\frac{1}{N}\sum_{i=1}^{N}t_{i}\right) \int
dQ\delta \left( Q- \frac{1}{N}\sum_{i=1}^{N}D_{i}t_{i}\right)
\times
\end{equation*}
\begin{equation}
\exp \left\{ \beta N\left[ Q+\frac{\sigma ^{2}}{2k\phi }\right] \right\} .
\end{equation}
We now introduce integral representations of the $\delta $-functions. In the
thermodynamic limit, we invoke the law of large numbers in order to see that
the final expression for the free energy is self-averaging with respect to
disorder. Thus, we have
\begin{equation}
Y\left\{ D_{i}\right\} =Y=\int d\phi \int dQ\exp \left[ -\beta
N\mathcal{G} \right] ,
\end{equation}
with
\begin{equation}
\mathcal{G}=-Q-\frac{\sigma ^{2}}{2k\phi }-\frac{1}{\beta N}\ln
\Omega ,
\end{equation}
where
\begin{equation}
\Omega =\int \int dxdy\exp \left[ N \mathcal{F} \left( x,y\right)
\right] ,
\end{equation}
\begin{equation}
\mathcal{F} \left( x,y\right) =\phi x+Qy+\left\langle \ln \left[
1+\exp \left( -x-yD\right) \right] \right\rangle ,
\end{equation}
and $\left\langle ...\right\rangle $ indicates an average with respect to
the probability distribution $p\left( D\right) $. A saddle-point calculation
leads to an asymptotic expression of $\Omega $, which in turn leads to the
final (Gibbs)\ free energy of this system. For simple forms of the
probability distribution we can even write some analytic expressions.

\subsection{Bimodal distribution of disorder}

\label{sec:bimodal}For a bimodal distribution, given by Eq.
(\ref{bimodal}), we can write analytical expressions. In
particular, with $D_{1}=0$ and $ D_{2}=D>0$, and defining
\begin{equation}
\chi =\frac{Q}{D},
\end{equation}
we have
\begin{equation}
\label{gibbs} \mathcal{G}=\mathcal{G}\left(
T,\sigma;\phi,\chi\right) =-D\chi -\frac{\sigma ^{2}}{2k\phi }-
\frac{1}{\beta }s\left( \phi ,\chi \right) ,
\end{equation}
where
\begin{equation*}
s\left( \phi ,\chi \right) = - \left[(1-c)+ \chi-\phi \right] \ln
\left[ (1-c) + \chi-\phi \right]+(\chi -\phi) \ln (\chi -\phi) +
\end{equation*}
\begin{equation}
 +\chi
\ln \chi - (c-\chi) \ln (c- \chi)+ c \ln c + (1-c) \ln (1-c).
\end{equation}
In Fig. \ref{fig5}, we draw equilibrium values of $\chi $ and
$\phi $ for $ c=0.7$ and applied stress $\sigma =0.3$. It is
interesting to see that there are two distinct values of $\chi $
for each value of $\phi $, corresponding to different
temperatures. This implies that there are two different
microscopic states at each point of the load curve $(\sigma
,\epsilon )$. The turning point around $T=0.1$ corresponds to the
extremum of $\phi $ in Fig. \ref{fig6}. Figure \ref{fig7} reports
the dependence of $\phi$ on $\sigma$ for different temperatures.
In the next Section we make some additional comments on these
results.

\subsection{Uniform disorder}

Let us look at the case of a uniform distribution of dissociation
energies in the interval $[a,b]$. Since we can no longer express
the free energy in terms of elementary functions, we are forced to
resort to some numerical calculations. In Fig. \ref{fig8}, we
compare the phase diagrams of the disordered DTFB and the uniform
TFB models. The solid line represents the spinodal for the
Helmholtz free energy. In the DTFB model, the averages are taken
with respect to two different uniform distributions. It is seen
that the main effect of disorder is the weakening of the system at
low temperature. In the case of wider distributions of disorder,
the spinodal displays a plateau as $T \rightarrow 0$, implying a
constant failure stress in that region. For temperature larger
than $1$, the TFB and the DTFB models display identical behavior,
and the disorder does not affect the system.

\section{The Edwards-Anderson parameter}

In the absence of disorder we may take $\phi $ as the only order
parameter of the system. We may replace the ensemble average of a
quantity  $\langle A \rangle$ by a time average $\lim_{T
\rightarrow \infty} 1/T \, \int A(t)dt $, and thus interpret the
average fraction of intact fibers as the average fraction of time
during which a fiber remains intact, corresponding to the thermal
average $\langle t_{i} \rangle_T$. This is the same for all the
fibers since all of them are identical. If disorder is introduced,
and the dissociation energy of each fiber becomes different, this
no longer holds. The thermal average may be different for each
fiber, depending on the value of $D$. Thus, the same macroscopic
state, $\phi $, may correspond to different microscopic states,
and it might be interesting to distinguish among them, even for
practical purposes. A quantity which can operate this distinction
is an analog of the Edwards-Anderson parameter \cite{edwards75},
\begin{equation*}
q=\langle \langle t_{i} \rangle_{T}^2 \rangle.
\end{equation*}

At constant strain, we can write
\begin{equation}
q(\varepsilon ,T)=\int dDp(D)\frac{1}{\left[    1+\exp \left(
\frac{1}{2}\beta k\varepsilon ^{2}-\beta D\right) \right] ^{2}},
\label{genq}
\end{equation}
from which it follows that
\begin{equation*}
\lim_{T\rightarrow \infty }q=\lim_{T\rightarrow \infty }\phi
^{2}=\frac{1}{4},
\end{equation*}
as it could have been anticipated. Then, it seems useful to
introduce the quantity $r=q-\phi ^{2}$ as a measure of the degree
of \textquotedblleft freezing\textquotedblright\ of the system.

Figure \ref{fig9} shows a detail of the strain stress curve of
Fig. \ref{fig3} for the bimodal distribution at different
temperatures. This plot shows that the curves for $T=0.1$ and
$T=0.3$ cross at a point in the $(\varepsilon ,\sigma )$ plane.
This same macroscopic state corresponds to different underlying
microscopic states characterized by different values of the
elastic modulus $c$. These values can be related to the
microscopic state by the analog of the Edwards-Anderson parameter.
In fact we have:
\begin{equation}
c=\frac{\partial \sigma }{\partial \epsilon }=k\phi
-k\frac{\epsilon ^{2}}{T} (\phi -q).
\end{equation}
From this expression, together with $\sigma =k\phi \epsilon $, we
obtain the values of $\phi $ and $q$. In particular, at the
failure, $c=0$, we have
\begin{equation}
q=\frac{\sigma }{k\varepsilon }\left[ 1-\frac{T}{k\varepsilon ^{2}}\right] .
\end{equation}

Another instance of this lack of uniqueness is given in Figs.
\ref{fig10} a) and  b), which have been generated by simulated
annealing. The graphs in these figures show the dependence of the
``degree of freezing'' $r=q-\phi^2$ on the applied stress at some
low temperature values. The corresponding curves for $\phi$ are
given in Figs. \ref{fig7} a) and b). In analogy with the stress vs
strain diagram, the curves for different temperatures cross at
some points, where however $r$ assumes distinct values.

\section{Conclusions}

We have shown that the introduction of disorder in the
Thermodynamical Fiber Bundle model \cite{selinger91} affects its
behavior in many respects. Besides the onset of some expected
features, as a decrease in the failure stress at low temperatures,
there appear some new features associated with the distribution of
disorder. For a bimodal disorder distribution, which leads to an
analytically tractable problem, the fraction of integer fibers is
non-monotonic in terms of temperature, with an extremum at low
temperatures. As a consequence, at either constant stress or
constant strain, the system may display the same values of the
fraction of integer fibers at different temperatures. However, the
corresponding microscopic states are different, which is relevant
for understanding delayed fractures, and can be characterized by
different values of an analog of the Edwards Anderson parameter.

A.V. is grateful to Fergal Dalton for many useful discussions and
suggestions.  A.P. S.R.S acknowledge financial support from CNPq
and CNR.

\bibliographystyle{apsrev}
\bibliography{FBM-refs,FRAC-refs,TF-refs}

\begin{thebibliography}{23}
\expandafter\ifx\csname natexlab\endcsname\relax\def\natexlab#1{#1}\fi
\expandafter\ifx\csname bibnamefont\endcsname\relax
  \def\bibnamefont#1{#1}\fi
\expandafter\ifx\csname bibfnamefont\endcsname\relax
  \def\bibfnamefont#1{#1}\fi
\expandafter\ifx\csname citenamefont\endcsname\relax
  \def\citenamefont#1{#1}\fi
\expandafter\ifx\csname url\endcsname\relax
  \def\url#1{\texttt{#1}}\fi
\expandafter\ifx\csname urlprefix\endcsname\relax\def\urlprefix{URL }\fi
\providecommand{\bibinfo}[2]{#2}
\providecommand{\eprint}[2][]{\url{#2}}

\bibitem[{\citenamefont{Hermann and Roux}(1990)}]{roux90}
\bibinfo{author}{\bibfnamefont{H.}~\bibnamefont{Hermann}} \bibnamefont{and}
  \bibinfo{author}{\bibfnamefont{S.}~\bibnamefont{Roux}},
  \emph{\bibinfo{title}{Models for the fracture of disordered media}}
  (\bibinfo{publisher}{North Holland}, \bibinfo{year}{1990}).

\bibitem[{\citenamefont{Chakrabarti and Benguigui}(1997)}]{chakrabarti97}
\bibinfo{author}{\bibfnamefont{B.}~\bibnamefont{Chakrabarti}} \bibnamefont{and}
  \bibinfo{author}{\bibfnamefont{L.~G.} \bibnamefont{Benguigui}},
  \emph{\bibinfo{title}{Statistical physics of fracture and breakdown in
  disordered systems}} (\bibinfo{publisher}{Oxford Science Publications},
  \bibinfo{year}{1997}).

\bibitem[{\citenamefont{Alava et~al.}(2006)\citenamefont{Alava, Nukala, and
  Zapperi}}]{alava05}
\bibinfo{author}{\bibfnamefont{M.~J.} \bibnamefont{Alava}},
  \bibinfo{author}{\bibfnamefont{P.~K. V.~V.} \bibnamefont{Nukala}},
  \bibnamefont{and} \bibinfo{author}{\bibfnamefont{S.}~\bibnamefont{Zapperi}},
  \bibinfo{journal}{Advances in Physics} \textbf{\bibinfo{volume}{55}},
  \bibinfo{pages}{349} (\bibinfo{year}{2006}).

\bibitem[{\citenamefont{Peirce}(1926)}]{peirce26}
\bibinfo{author}{\bibfnamefont{F.~T.} \bibnamefont{Peirce}},
  \bibinfo{journal}{J. Textile Industry} \textbf{\bibinfo{volume}{17}},
  \bibinfo{pages}{355} (\bibinfo{year}{1926}).

\bibitem[{\citenamefont{Daniels}(1945)}]{daniels45}
\bibinfo{author}{\bibfnamefont{H.~E.} \bibnamefont{Daniels}},
  \bibinfo{journal}{Proc. R. Soc. A} \textbf{\bibinfo{volume}{183}},
  \bibinfo{pages}{405} (\bibinfo{year}{1945}).

\bibitem[{\citenamefont{Hansen and Hemmer}(1994)}]{hansen94}
\bibinfo{author}{\bibfnamefont{A.}~\bibnamefont{Hansen}} \bibnamefont{and}
  \bibinfo{author}{\bibfnamefont{P.~C.} \bibnamefont{Hemmer}},
  \bibinfo{journal}{Phys. Lett. A} \textbf{\bibinfo{volume}{184}},
  \bibinfo{pages}{394} (\bibinfo{year}{1994}).

\bibitem[{\citenamefont{Petri et~al.}(1994)\citenamefont{Petri, Paparo,
  Vespignani, Alippi, and Costantini}}]{petri94}
\bibinfo{author}{\bibfnamefont{A.}~\bibnamefont{Petri}},
  \bibinfo{author}{\bibfnamefont{G.}~\bibnamefont{Paparo}},
  \bibinfo{author}{\bibfnamefont{A.}~\bibnamefont{Vespignani}},
  \bibinfo{author}{\bibfnamefont{A.}~\bibnamefont{Alippi}}, \bibnamefont{and}
  \bibinfo{author}{\bibfnamefont{M.}~\bibnamefont{Costantini}},
  \bibinfo{journal}{Phys. Rev. Lett.} \textbf{\bibinfo{volume}{73}},
  \bibinfo{pages}{3423} (\bibinfo{year}{1994}).

\bibitem[{\citenamefont{Caldarelli et~al.}(1996)\citenamefont{Caldarelli,
  DiTolla, and Petri}}]{caldarelli96}
\bibinfo{author}{\bibfnamefont{G.}~\bibnamefont{Caldarelli}},
  \bibinfo{author}{\bibfnamefont{F.}~\bibnamefont{DiTolla}}, \bibnamefont{and}
  \bibinfo{author}{\bibfnamefont{A.}~\bibnamefont{Petri}},
  \bibinfo{journal}{Phys. Rev. Lett.} \textbf{\bibinfo{volume}{77}},
  \bibinfo{pages}{2503} (\bibinfo{year}{1996}).

\bibitem[{\citenamefont{Zapperi et~al.}(1999)\citenamefont{Zapperi, Ray,
  Stanley, and Vespignani}}]{zapperi99}
\bibinfo{author}{\bibfnamefont{S.}~\bibnamefont{Zapperi}},
  \bibinfo{author}{\bibfnamefont{P.}~\bibnamefont{Ray}},
  \bibinfo{author}{\bibfnamefont{H.~E.} \bibnamefont{Stanley}},
  \bibnamefont{and}
  \bibinfo{author}{\bibfnamefont{A.}~\bibnamefont{Vespignani}},
  \bibinfo{journal}{Physical Review E} \textbf{\bibinfo{volume}{59}},
  \bibinfo{pages}{5049} (\bibinfo{year}{1999}).

\bibitem[{\citenamefont{Shahidzadeh-Bonn
  et~al.}(2005)\citenamefont{Shahidzadeh-Bonn, Vie´, Chateau, Roux, and
  Bonn}}]{shahidzadeh05}
\bibinfo{author}{\bibfnamefont{N.}~\bibnamefont{Shahidzadeh-Bonn}},
  \bibinfo{author}{\bibfnamefont{P.}~\bibnamefont{Vie´}},
  \bibinfo{author}{\bibfnamefont{X.}~\bibnamefont{Chateau}},
  \bibinfo{author}{\bibfnamefont{J.-N.} \bibnamefont{Roux}}, \bibnamefont{and}
  \bibinfo{author}{\bibfnamefont{D.}~\bibnamefont{Bonn}},
  \bibinfo{journal}{Phys. Rev. Lett.} \textbf{\bibinfo{volume}{95}},
  \bibinfo{pages}{175501} (\bibinfo{year}{2005}).

\bibitem[{\citenamefont{Golubovi\'c and Feng}(1991)}]{golubovich91}
\bibinfo{author}{\bibfnamefont{L.}~\bibnamefont{Golubovi\'c}} \bibnamefont{and}
  \bibinfo{author}{\bibfnamefont{S.}~\bibnamefont{Feng}},
  \bibinfo{journal}{Physical Review A} \textbf{\bibinfo{volume}{43}},
  \bibinfo{pages}{5223} (\bibinfo{year}{1991}).

\bibitem[{\citenamefont{Pauchard and Meunier}(1993)}]{pauchard93}
\bibinfo{author}{\bibfnamefont{L.}~\bibnamefont{Pauchard}} \bibnamefont{and}
  \bibinfo{author}{\bibfnamefont{J.}~\bibnamefont{Meunier}},
  \bibinfo{journal}{Phys. Rev. Lett.} \textbf{\bibinfo{volume}{70}},
  \bibinfo{pages}{3565} (\bibinfo{year}{1993}).

\bibitem[{\citenamefont{Kitamura et~al.}(1997)\citenamefont{Kitamura, Maksimov,
  and Nishioka}}]{kitamura97}
\bibinfo{author}{\bibfnamefont{K.}~\bibnamefont{Kitamura}},
  \bibinfo{author}{\bibfnamefont{I.}~\bibnamefont{Maksimov}}, \bibnamefont{and}
  \bibinfo{author}{\bibfnamefont{K.}~\bibnamefont{Nishioka}},
  \bibinfo{journal}{Philosophical Magazine Letters}
  \textbf{\bibinfo{volume}{75}}, \bibinfo{pages}{343} (\bibinfo{year}{1997}).

\bibitem[{\citenamefont{Maksimov et~al.}(2001)\citenamefont{Maksimov, Kitamura,
  and Nishioka}}]{kitamura01}
\bibinfo{author}{\bibfnamefont{I.}~\bibnamefont{Maksimov}},
  \bibinfo{author}{\bibfnamefont{K.}~\bibnamefont{Kitamura}}, \bibnamefont{and}
  \bibinfo{author}{\bibfnamefont{K.}~\bibnamefont{Nishioka}},
  \bibinfo{journal}{Philosophical Magazine Letters}
  \textbf{\bibinfo{volume}{81}}, \bibinfo{pages}{547} (\bibinfo{year}{2001}).

\bibitem[{\citenamefont{Bonn et~al.}(1998)\citenamefont{Bonn, Kellay, Prochnow,
  Ben-Djemiaa, and Meunier}}]{bonn98}
\bibinfo{author}{\bibfnamefont{D.}~\bibnamefont{Bonn}},
  \bibinfo{author}{\bibfnamefont{H.}~\bibnamefont{Kellay}},
  \bibinfo{author}{\bibfnamefont{M.}~\bibnamefont{Prochnow}},
  \bibinfo{author}{\bibfnamefont{K.}~\bibnamefont{Ben-Djemiaa}},
  \bibnamefont{and} \bibinfo{author}{\bibfnamefont{J.}~\bibnamefont{Meunier}},
  \bibinfo{journal}{Science} \textbf{\bibinfo{volume}{280}},
  \bibinfo{pages}{265} (\bibinfo{year}{1998}).

\bibitem[{\citenamefont{Politi et~al.}(2002)\citenamefont{Politi, Ciliberto,
  and Scorretti}}]{politi02}
\bibinfo{author}{\bibfnamefont{A.}~\bibnamefont{Politi}},
  \bibinfo{author}{\bibfnamefont{S.}~\bibnamefont{Ciliberto}},
  \bibnamefont{and}
  \bibinfo{author}{\bibfnamefont{R.}~\bibnamefont{Scorretti}},
  \bibinfo{journal}{Physical Review E} \textbf{\bibinfo{volume}{66}},
  \bibinfo{pages}{026107} (\bibinfo{year}{2002}).

\bibitem[{\citenamefont{Santucci et~al.}(2003)\citenamefont{Santucci, Vanel,
  Guarino, Scorretti, and Ciliberto}}]{santucci03}
\bibinfo{author}{\bibfnamefont{S.}~\bibnamefont{Santucci}},
  \bibinfo{author}{\bibfnamefont{L.}~\bibnamefont{Vanel}},
  \bibinfo{author}{\bibfnamefont{A.}~\bibnamefont{Guarino}},
  \bibinfo{author}{\bibfnamefont{R.}~\bibnamefont{Scorretti}},
  \bibnamefont{and}
  \bibinfo{author}{\bibfnamefont{S.}~\bibnamefont{Ciliberto}},
  \bibinfo{journal}{Europhys. Lett.} \textbf{\bibinfo{volume}{62}},
  \bibinfo{pages}{320} (\bibinfo{year}{2003}).

\bibitem[{\citenamefont{Santucci et~al.}(2004)\citenamefont{Santucci, Vanel,
  and Ciliberto}}]{santucci04}
\bibinfo{author}{\bibfnamefont{S.}~\bibnamefont{Santucci}},
  \bibinfo{author}{\bibfnamefont{L.}~\bibnamefont{Vanel}}, \bibnamefont{and}
  \bibinfo{author}{\bibfnamefont{S.}~\bibnamefont{Ciliberto}},
  \bibinfo{journal}{Physical Review Letters} \textbf{\bibinfo{volume}{93}},
  \bibinfo{pages}{95505} (\bibinfo{year}{2004}).

\bibitem[{\citenamefont{Sornette and Ouillon}(2005)}]{sornette05}
\bibinfo{author}{\bibfnamefont{D.}~\bibnamefont{Sornette}} \bibnamefont{and}
  \bibinfo{author}{\bibfnamefont{G.}~\bibnamefont{Ouillon}},
  \bibinfo{journal}{Physical Review Letters} \textbf{\bibinfo{volume}{94}},
  \bibinfo{eid}{038501} (\bibinfo{year}{2005}).

\bibitem[{\citenamefont{Selinger et~al.}(1991)\citenamefont{Selinger, Wang,
  Gelbart, and Ben-Shaul}}]{selinger91}
\bibinfo{author}{\bibfnamefont{R.~L.~B.} \bibnamefont{Selinger}},
  \bibinfo{author}{\bibfnamefont{Z.-G.} \bibnamefont{Wang}},
  \bibinfo{author}{\bibfnamefont{W.~M.} \bibnamefont{Gelbart}},
  \bibnamefont{and}
  \bibinfo{author}{\bibfnamefont{A.}~\bibnamefont{Ben-Shaul}},
  \bibinfo{journal}{Phys. Rev. A.} \textbf{\bibinfo{volume}{43}},
  \bibinfo{pages}{4396} (\bibinfo{year}{1991}).

\bibitem[{\citenamefont{Brenner}(1962)}]{brenner62}
\bibinfo{author}{\bibfnamefont{S.~S.} \bibnamefont{Brenner}},
  \bibinfo{journal}{Journal of Applied Physics} \textbf{\bibinfo{volume}{33}},
  \bibinfo{pages}{33} (\bibinfo{year}{1962}).

\bibitem[{\citenamefont{Virgilii et~al.}()\citenamefont{Virgilii, Petri, and
  Salinas}}]{next}
\bibinfo{author}{\bibfnamefont{A.}~\bibnamefont{Virgilii}},
  \bibinfo{author}{\bibfnamefont{A.}~\bibnamefont{Petri}}, \bibnamefont{and}
  \bibinfo{author}{\bibfnamefont{S.~R.} \bibnamefont{Salinas}},
  \bibinfo{note}{in preparation}.

\bibitem[{\citenamefont{Edwards and Anderson}(1975)}]{edwards75}
\bibinfo{author}{\bibfnamefont{S.}~\bibnamefont{Edwards}} \bibnamefont{and}
  \bibinfo{author}{\bibfnamefont{P.}~\bibnamefont{Anderson}},
  \bibinfo{journal}{J. Phys. F: Metal Phys.} p. \bibinfo{pages}{965}
  (\bibinfo{year}{1975}).

\end{thebibliography}

\begin{figure}
\epsfig{file=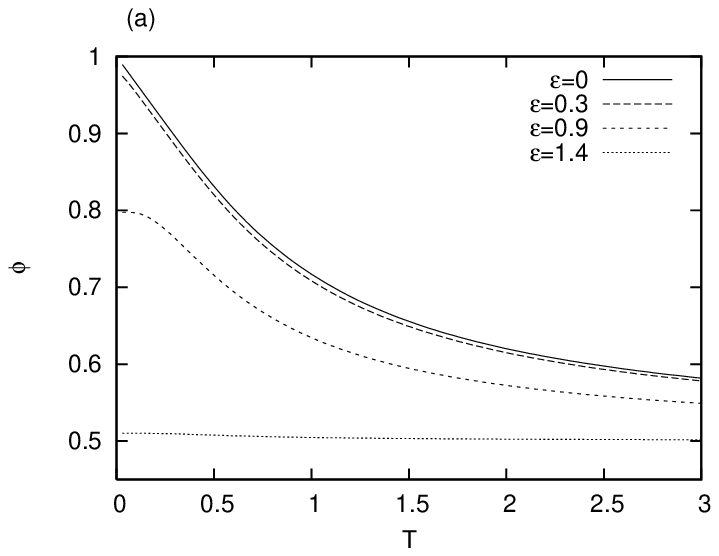, width=8cm, height=6cm}
\epsfig{file=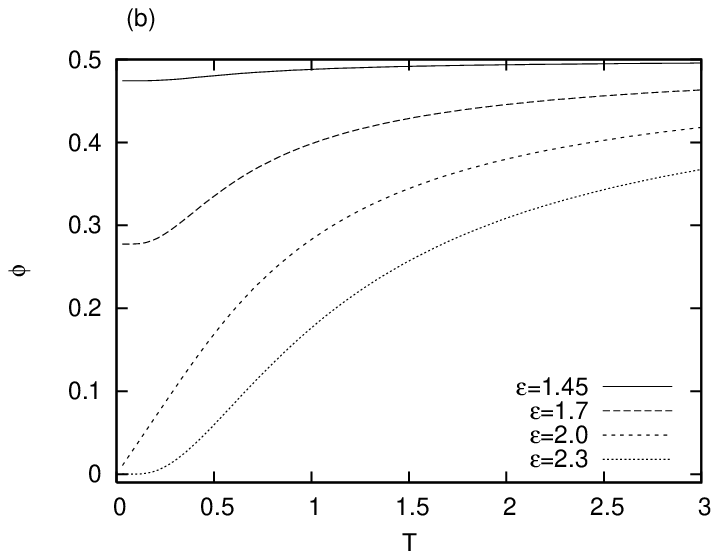, width=8cm, height=6cm}
\epsfig{file=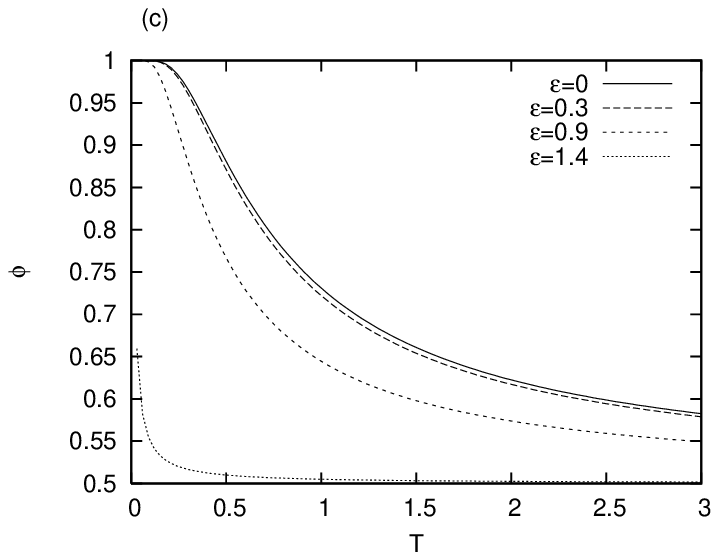, width=8cm, height=6cm}
\epsfig{file=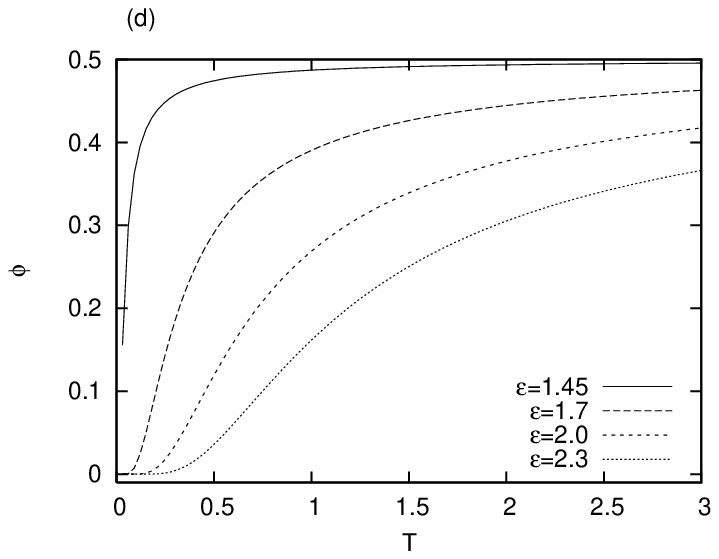, width=8cm, height=6cm} \caption{Fraction
of integer fibers $\phi$ as a function of temperature for
different values of the applied strain $\varepsilon$: a), b) in
the DTFB model with a uniform distribution of disorder in $[0,2]$;
c), d) in the homogeneous TFB model (with no disorder) with $D=1$.
For $0.9 \approx \varepsilon \approx \sqrt{2}$ these curves differ
very little from each other. We always take $k=1$.} \label{fig1}
\end{figure}

\begin{figure}
 \epsfig{file=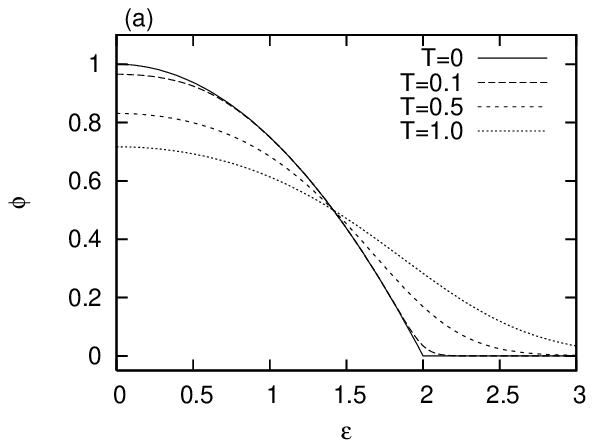, width=8cm, height=6cm}
\epsfig{file=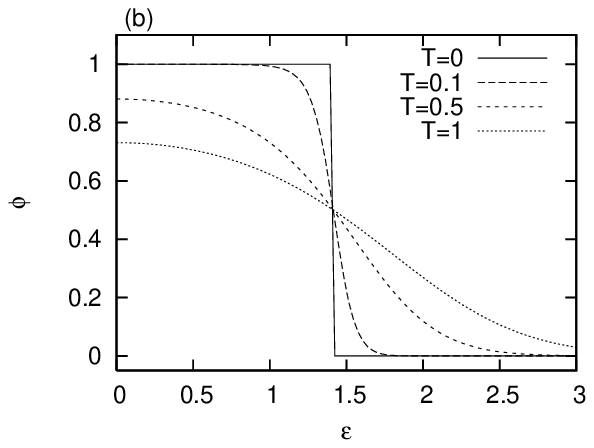, width=8cm, height=6cm} \caption{Fraction
of integer fibers $\phi$ as a function of strain at different
temperatures: a) in the DTFB model with a homogeneous distribution
of disorder with limits $0$ and $2$;  b)  in the homogeneous TFB
model (no disorder) with $D=1$. We always take $k=1$.}
\label{fig2}
\end{figure}

\begin{figure}
 \epsfig{file=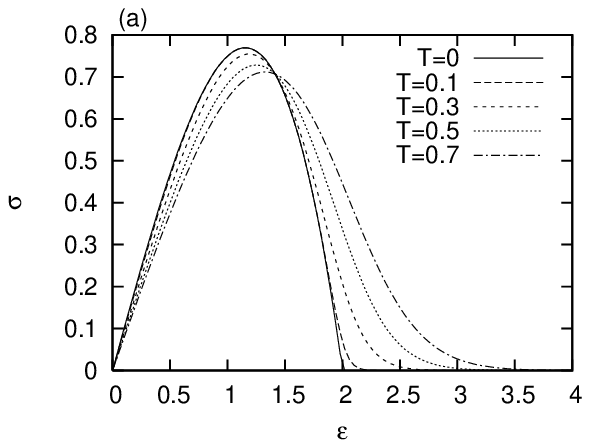, width=5cm, height=3cm}
\epsfig{file=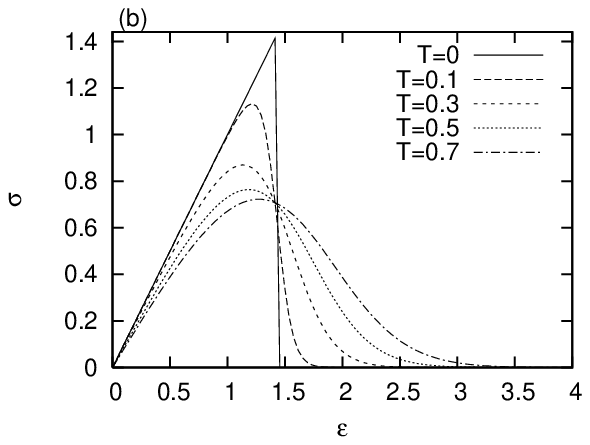, width=5cm, height=3cm}
 \epsfig{file=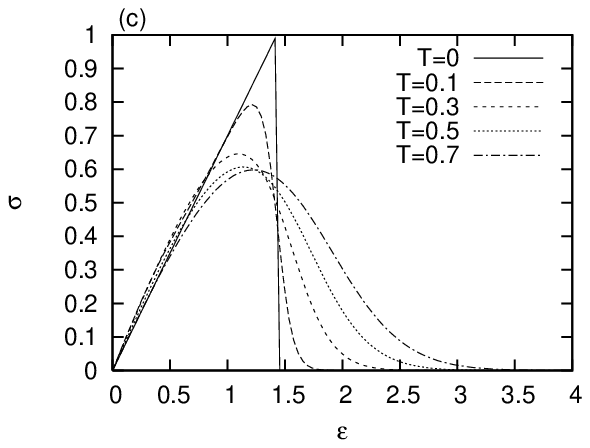, width=5cm, height=3cm}
 \caption{Stress-strain characteristic curves
at different temperatures: a) disordered DTFB model with a
homogeneous distribution of disorder in $[0,2]$;  b) homogeneous
TFB model (no disorder) with $D=1$; the case $T=0$ corresponds to
the ordinary FB model; c) DTFB model for bimodal disorder,
Eq.~(\protect{\ref{bimodal}}), with parameters $c=0.7$, $D_1=0$
and $D_2=1$. We always take $k=1$.} \label{fig3}
\end{figure}

\begin{figure}
 \epsfig{file=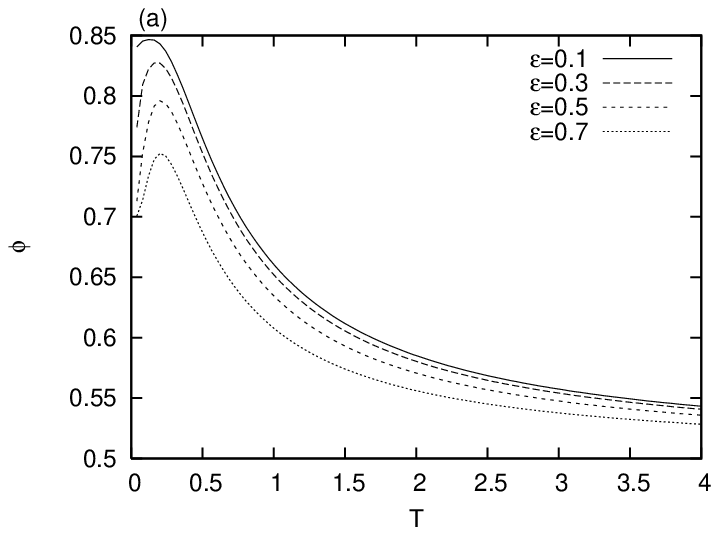, width=8cm, height=6cm}
\epsfig{file=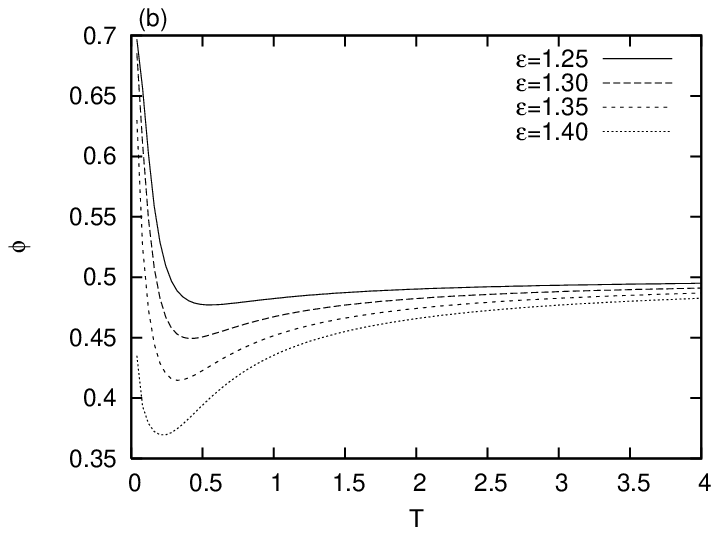, width=8cm, height=6 cm}
 \caption{Fraction of integer fibers $\phi$ versus temperature $T$ for a bimodal
 distribution, Eq.~(\protect{\ref{bimodal}}), with
$D_1=0$, $D_2=1$, $c=0.7$ at constant strain with $k=1$.}
\label{fig4}
\end{figure}

\begin{figure}
 \epsfig{file=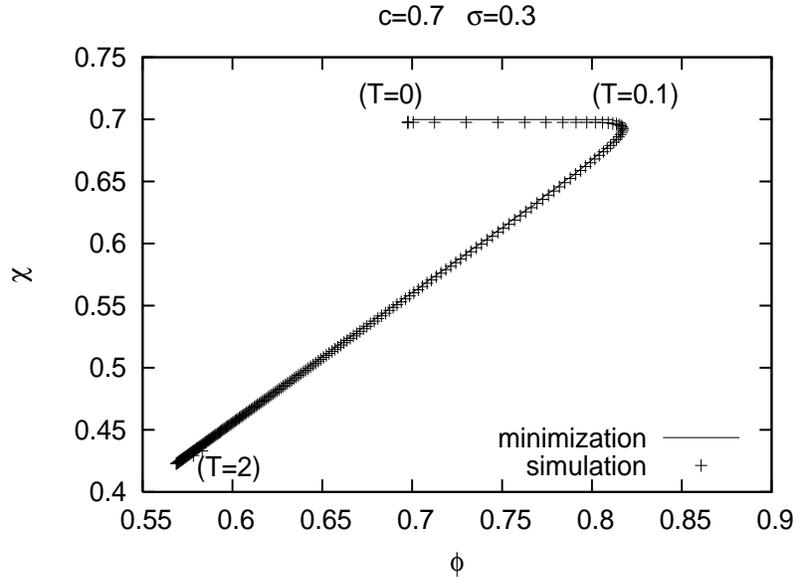, width=11cm, height=8cm}
 \caption{Parameters $\chi$ vs $\phi$, after minimization of
 Eq.~(\protect{\ref{gibbs}}) and from numerical
simulations, in the case of bimodal disorder,
Eq.~(\protect{\ref{bimodal}}), with $c=0.7$ and stress
$\sigma=0.3$,  from Eq.(\protect{\ref{gibbs}}), and $k=1$.  It is
seen that $\chi$ assumes two different values for the same value
of $\phi$.} \label{fig5}
\end{figure}

\begin{figure}
 \epsfig{file=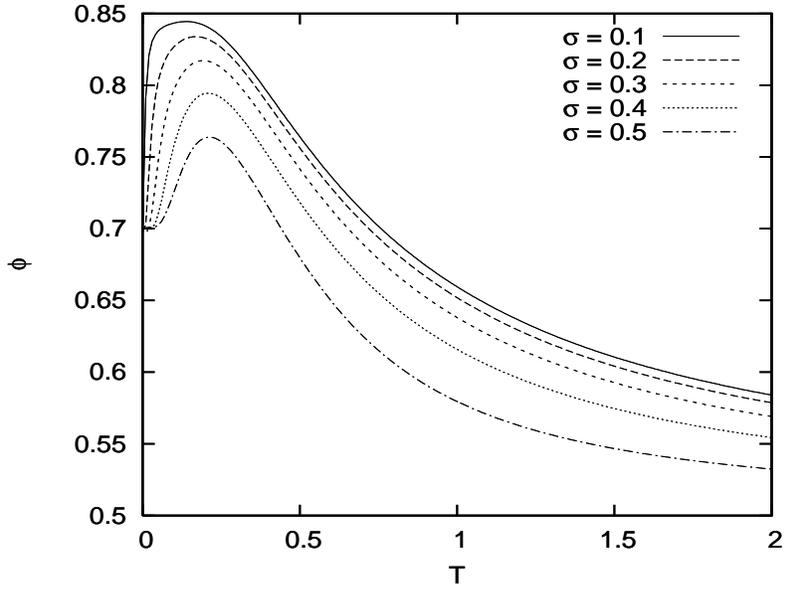, width=11cm, height=8cm}
 \caption{Plot of  $\phi$ for a bimodal
 disorder at constant stress, Eq.~(\protect{\ref{bimodal}}), with
$D_1=0$, $D_2=1$, $c=0.7$, and a  Young modulus $k=1$.}
\label{fig6}
\end{figure}

\begin{figure}
 \epsfig{file=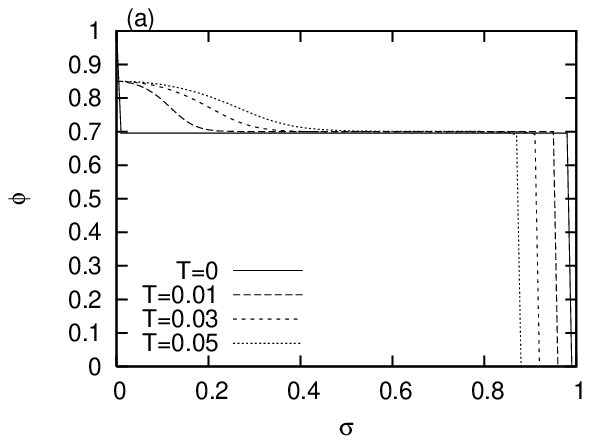, width=8cm, height=6cm}
 \epsfig{file=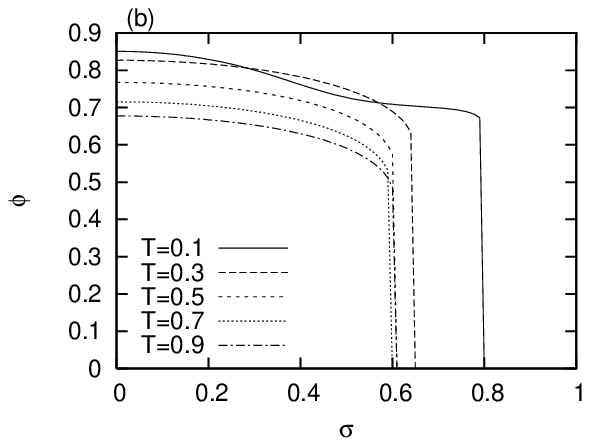, width=8cm, height=6cm}
 \caption{Stress dependence of $\phi$ at various
 temperatures for the bimodal distribution of disorder with
$D_1=0$, $D_2=1$, $c=0.7$, and a Young modulus $k=1$. }
\label{fig7}
\end{figure}

\begin{figure}
 \epsfig{file=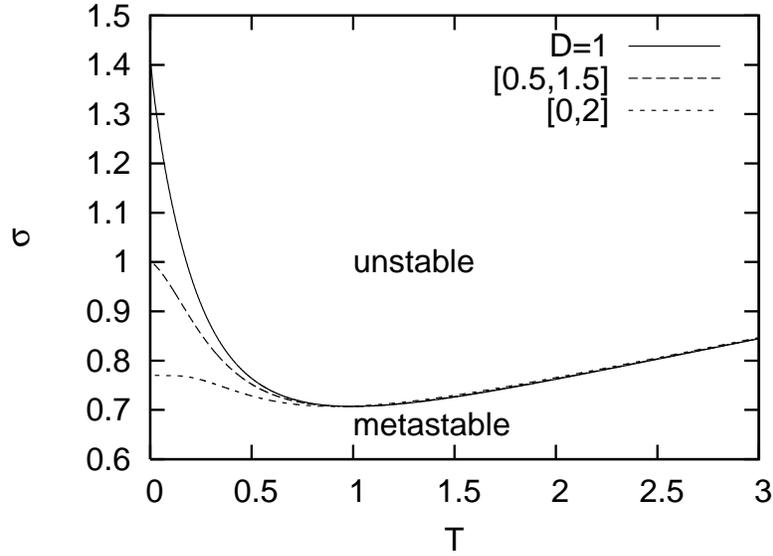, width=11cm}
\caption{Comparison between the phase diagram of a case without
disorder (TFB with D=1) and the phase diagrams for two
homogeneously disordered cases: $D_i \in [0.5; 1.5]$ and $D_i \in
[0; 2]$. The Young modulus is $k=1$.} \label{fig8}
\end{figure}

\begin{figure}
 \epsfig{file=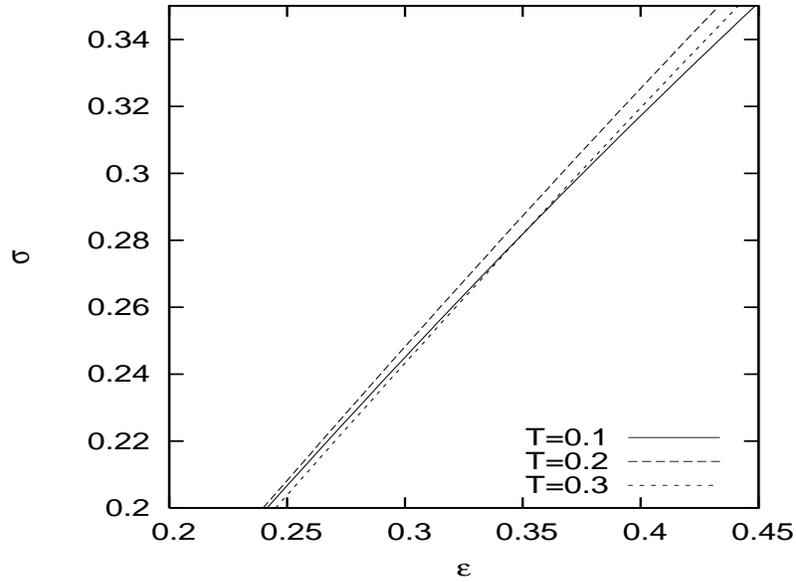, width=11cm,
height=8cm}  \caption{A bimodal distribution of disorder with
$D_1=0$, $D_2=1$, $c=0.7$ (we always take $k=1$). It is seen that
for two different temperatures, $T=0.1$ and $T=0.3$, stress and
strain assume the same values (around $\epsilon=0.35$ and
$\sigma=0.28$). The  different microscopic configurations can be
distinguished  by different values of the Edwards-Anderson
parameter $q$.}
\label{fig9}
\end{figure}

\begin{figure}
 \epsfig{file=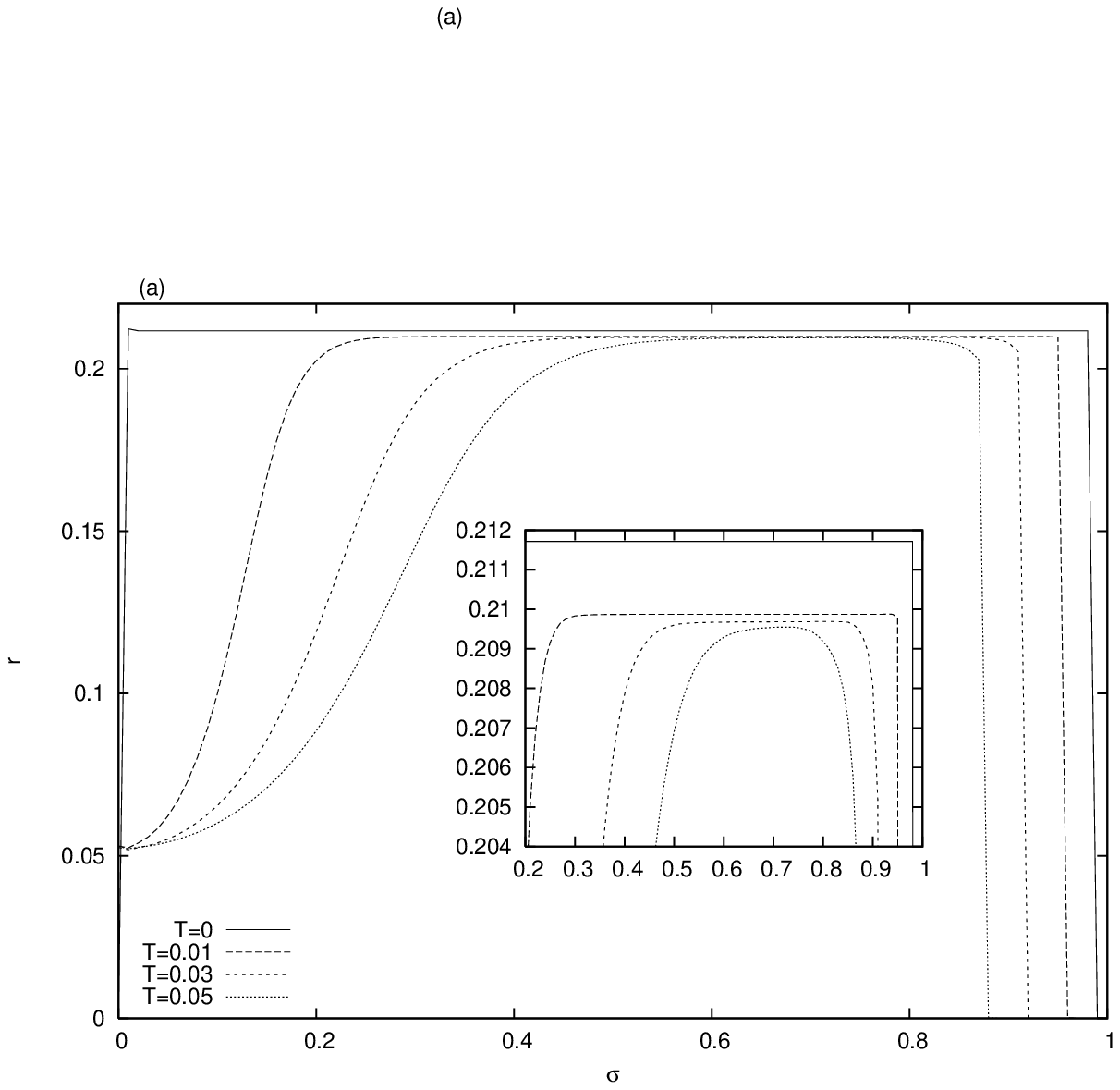, width=8cm, height=6cm}
 \epsfig{file=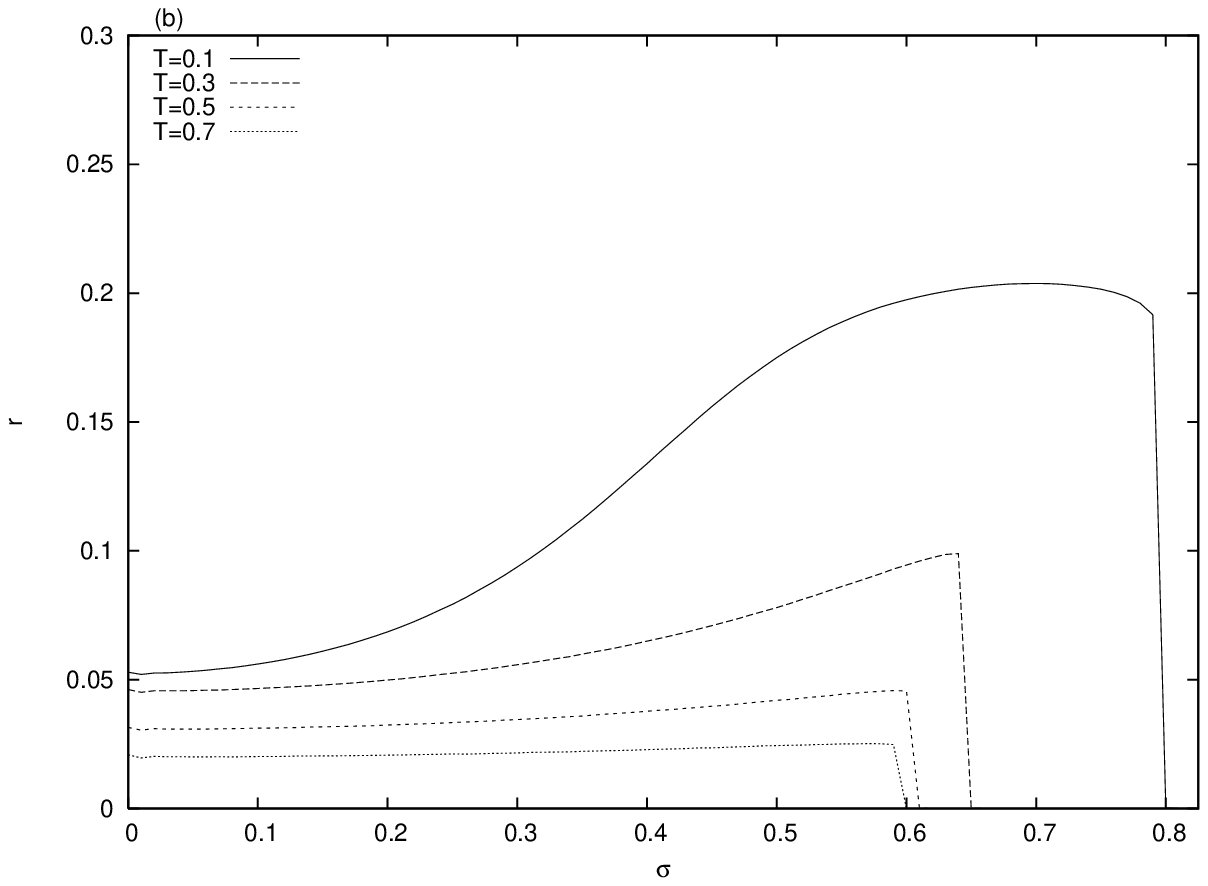, width=8cm, height=6cm}
 \caption{Stress dependence of $r$ at various
 temperatures for the bimodal distribution of disorder with
$D_1=0$, $D_2=1$, $c=0.7$, and a Young modulus $k=1$. The inset in
a) shows that curves for distinct temperatures have distinct
values and do not intersect.}
 \label{fig10}
 \end{figure}

\end{document}